\documentclass{article}
\usepackage{arxiv}
\usepackage[utf8]{inputenc} 
\usepackage[T1]{fontenc}
\usepackage{algorithm}
\usepackage{amsmath,amssymb,amsfonts}
\usepackage{algorithmic}
\usepackage{graphicx}
\usepackage{textcomp}
\usepackage{subfigure}
\usepackage{subfig}
\usepackage{xcolor}
\usepackage{soul}
\usepackage{todonotes}
\usepackage{multirow}
\usepackage{multicol}
\usepackage{float}
\usepackage[belowskip=-6pt,aboveskip=0pt]{caption}
\usepackage[nolist]{acronym}

\begin{document}
\title{A Novel Malware Detection Mechanism based on Features Extracted from Converted \\ Malware Binary Images\\
}
\author{Abhijitt Dhavlle and Sanket Shukla \\
Electrical and Computer Engineering \\
\textit{George Mason University}\\
Fairfax, USA. \\
(adhavlle, sshukla4)@gmu.edu}

\maketitle

\begin{abstract}
Our computer systems since decades have been threatened by various types of harware and software attacks of which Malwares have been one of them. These malwares have the ability to steal, destroy, contaminate, gain unintended access or even disrupt the entire system. There have been techniques to detect malwares by performing static and dynamic analysis of malware files, but, stealthy malware has circumvented the static analysis method and for dynamic analysis there have been previous works that propose different methods to detect malwares but, in this work we propose a novel technique to detect malwares. We use malware binary images and then extract different features from the same and then employ different ML-classifiers on the dataset thus obtained. We show that this technique is successful in differentiating classes of malware based on the features extracted. 

\end{abstract}

\section{Introduction}
\label{intro}
The hardware security discipline in recent years experienced a
plethora of threats like the Malware attacks \cite{Dhavlle_DATE'21, Dhavlle_ISCAS'21, Meraj_ISCAS'21, SMPD_DAC'19, Sanket_CASES'19, Sanket_ICMLA'19, Sanket_ICTAI'19}, Side-Channel Attacks \cite{Yarom_usenix_'14,Gruss_dimva_'16, Abhijitt_isqed'20, Dhavlle_TCAD'21}, Hardware Trojan attacks \cite{Meraj_AsianHOST'20}, reverse engineering threats \cite{Kolhe_GLSVLSI'19, Kolhe_ICCAD'19, Hassan_ISQED'20} and so on. We focus on the malware detection technique here along with some state-of-the-art works.
Malicious Software, generally known as `malware' is a software program developed by an attacker to gain unintended access of a computer system for performing unwanted actions and malicious activities like stealing data, sensitive information  ( like passwords, SSN's), contaminating and manipulating data without users consent. According to 2018 threat report by McAfee labs, about 73 million malicious files, 66 million malicious IP addresses and and 63 million malicious URL's were detected and declared as risky. Similar threat report by McAfee reported 57.6 million malicious files in 2017. This rapidly increasing trend of generation of malware is a serious threat and global concern for the community. It results in need to develop a promising and comprehensive malware detection methods with robustness. Traditional and primitive software based solutions for malware detection methods such as signature-based and semantics-based anomaly detection techniques induce remarkable computational overheads to the system \cite{sign_based_technique_1} \cite{sign_based_technique_2} \cite{sign_based_technique_3} \cite{sign_based_technique_4}.
Traditional approaches towards analyzing malware involve extraction of binary signatures from malware, comprising their fingerprint. There is an exponential increase in the number of new signatures released every year, due to the rapid escalation of malware.

Static code analysis and dynamic code analysis are some of the approaches used for analyzing malware. Static analysis looks for malicious patterns by disassembling the code and exploring the control flow of the executable. Whereas, in  dynamic analysis malicious code is executed in a virtual environment and based on the execution trace a behavioral report characterizing the executable is generated. These techniques have their pros and cons. Although, static analysis offers the most complete coverage but it suffers from code obfuscation. Prior to analysis, the executable has to be unpacked and decrypted, and even then, the analysis can be thwarted by problems of intractable complexity.
Dynamic analysis does not need the malware executable to be unpacked or decrypted and is also more efficient but it is time intensive and consumes resources, which results in scalibility issues. Moreover, sometimes environment does not satisfy the triggering conditions, leaving some malicious behaviors unobserved.

This paper, illustrates a completely different and a novel approach to characterize and analyze malware. Approach tends to represent a malware executable as a binary string of zeros and ones. Furthermore, this vector of binaries can be reshaped and converted into a matrix which can be later viewed as an image. Malware belonging to same family showed significant visual similarities in image texture. In section \ref{background} we discuss representing malware binaries as images. We consider malware classification problem as one of image classification problem. Existing classification techniques require either disassembly or execution whereas our method does not require either but still shows significant improvement in terms of performance. Moreover, our method is also resilient to popular obfuscation techniques such as section encryption. This automatic classification technique should be very valuable for anti-virus companies and security researchers who report thousands of malware everyday.

    
    
    




\section{Background and Motivation}
\label{background}

Several tools such as text editors and binary editors can both visualize and manipulate binary data. There have been several GUI-based tools which facilitate comparison of files. However, there has been limited research in visualizing malware. In \cite{5_Yoo_2004} Yoo used self organizing maps to detect and visualize malicious code inside an executable. In \cite{6_quist} Quist and Liebrock develop a visualization framework for reverse engineering. They identify functional areas and de-obfuscate through a node-link visualization where nodes represent the address and links represent state transitions between addresses. In \cite{7_trinius} Trinius et al. display the distributions of operations using treemaps and the sequence of operations using thread graphs. In \cite{8_Goodall} Goodall et al. develop a visual analysis environment that can aid software developers to understand the code better. They also show how vulnerabilities within software can be visualized in their environment.

While there hasn’t been much work on viewing malware as digital images, Conti et al. \cite{9_Conti} visualized raw binary data of primitive binary fragments such as text, C++ data structure, image data, audio data as images. In \cite{10_Conti} Conti et al. show that they can automatically classify the different binary fragments using statistical features. However, their analysis is only concerned with identifying primitive binary fragments and not malware. This work presents a similar approach by representing malware as grayscale images.

Several techniques have been proposed for clustering and classification of malware. These include both static analysis \cite{11_Karim05malwarephylogeny,12_Kolter,13_Gao,14_Tian,15_tian,16_Tian,17_Gheorghescu2006ANAV} as well as dynamic analysis \cite{18_park,19_Bailey:2007,20_Bayer_scalable,21_Rieck:2008}. We will review papers that specifically deal with classification of malware. In \cite{21_Rieck:2008} Rieck et al. used features based on behavioral analysis of malware to classify them according to their families. They used a labeled dataset of 10,072 malware samples labeled by an anti-virus software and divide the dataset into 14 malware families. Then they monitored the behavior of all the malware in a sandbox environment which generated a behavioral report. From the report, they generate a feature vector for every malware based on the frequency of some specific strings in the report. A Support Vector Machine is used for training and testing the feature on the 14 families and they report an average classification accuracy of 88\%.
In contrast to \cite{21_Rieck:2008}, Tian et al \cite{14_Tian} use a very simple feature, the length of a program, to classify 7 different types of Trojans and obtain an average accuracy of 88\%. However, their analysis was only done on 721 files. In \cite{15_tian,16_Tian} the same authors improve their above technique by using printable string information from the malware. They evaluated their method on 1521 malware consisting of 13 families and reported a classification accuracy of 98.8\%.
In \cite{18_park}, Park et al. classify malware based on detecting the maximal common sub graph in a behavioral graph. They demonstrate their results on a set of 300 malware in 6 families.

With respect to related works, our classification method does not require any disassembly or execution of the actual malware code. Moreover, the image textures used for classification provide more resilient features in terms of obfuscation techniques, and in particular for encryption. Finally, we evaluated our approach on a larger dataset consisting in 25 families within a malware corpus of 9,458 malware. The evaluation results show that our method offers similar precision at a lower computational cost.

\subsection{EDA Analysis of the Malign Dataset}

\begin{figure*}
    \centering
    \includegraphics[width=1\textwidth]{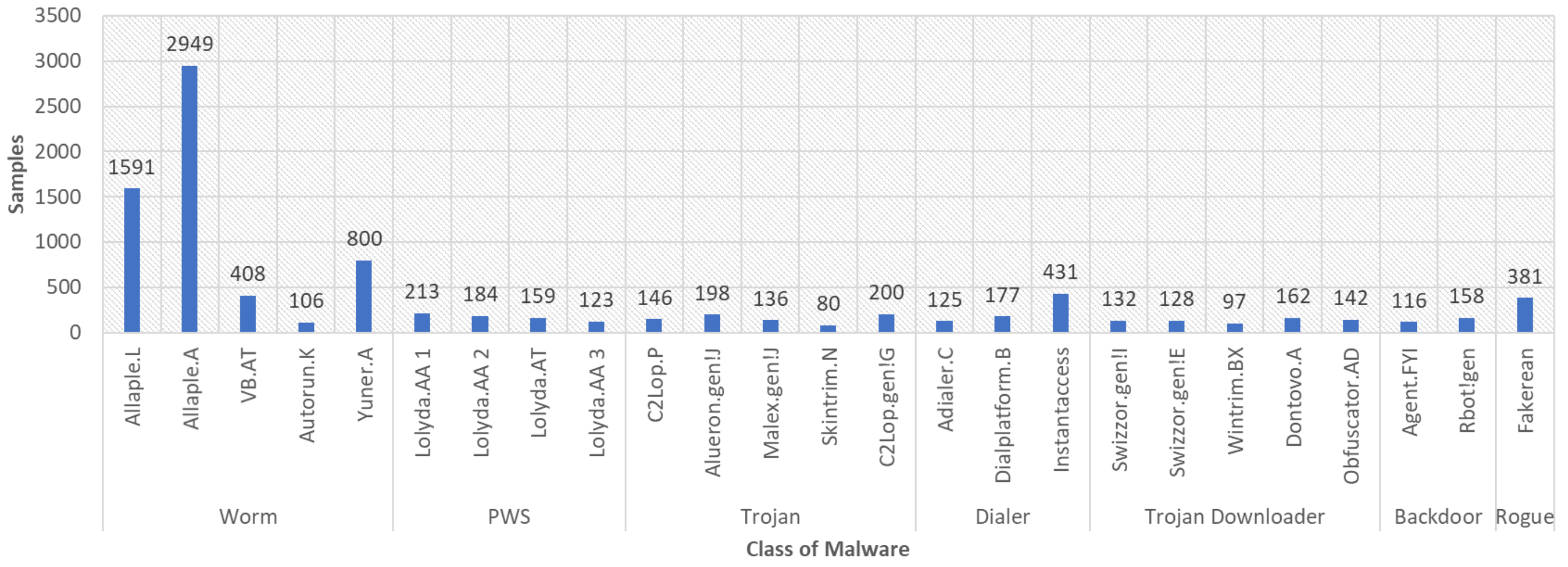}
    \vspace{10pt}
    \caption{Dataset distribution with Classes and Sub-Classes \vspace{10pt}}
    \vspace{1pt}
    \label{fig:malimg}
\end{figure*}

To mitigate the issue of classification of malware our first step was to perform exploratory data analysis (EDA) on the dataset. We are using malimg dataset for the analysis and the dataset distribution is as shown in Figure \ref{fig:malimg}. Prior to perform Exploratory data analysis we extracted some important features to create training and testing dataset. We extracted the following features for grayscale images in dataset: "energy, entropy, contrast, dissimilarity, homogeneity and correlation". "Entropy" defines statistical measure of randomness used to characterize the texture of the input image. "Energy" defines sum of squared elements in the gray level co-occurence matrix. "Contrast" defines intensity contrast between pixel and neighbor. "Dissimilarity" degree of dissimilarity between images. "Homogeneity" measures the closeness of the distribution of elements in the gray level co-occurence matrix to gray level co-occurence matrix  diagonal. "Correlation" defines correlation between a pixel and its neighbor over entire image.

\begin{figure}[!htb]
    \centering
    \includegraphics[width=0.45\textwidth]{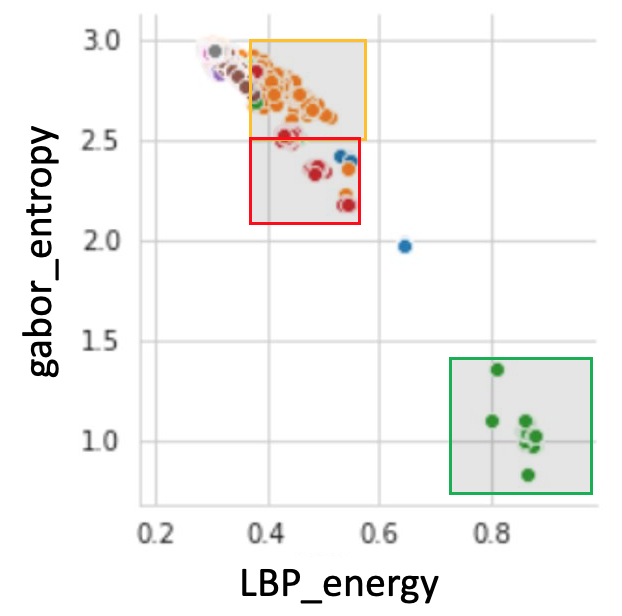}
    \caption{Gabor-entropy v/s LBP-energy}
    \label{fig:eda_1}
\end{figure}

\begin{figure}[!htb]
    \centering
    \includegraphics[width=0.45\textwidth]{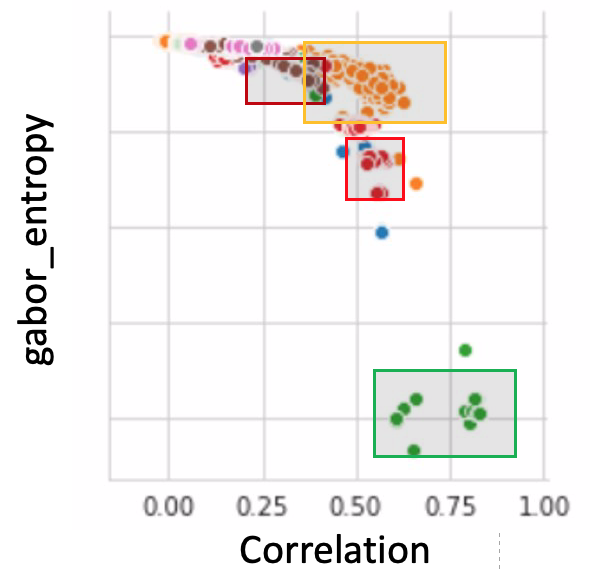}
    \caption{Gabor-entropy v/s Correlation}
    \label{fig:eda_2}
\end{figure}

Figure \ref{fig:eda_1} shows 2-D scatter plot of gabor-entropy v/s LBP-energy. Here we can easily cluster datapoints of 3 different classes, however there are many overlapping datapoints which makes it difficult to rely on these two features for malware classification. Similar when we select other features as shown in Figure \ref{fig:eda_2} gabor-entropy v/s correlation. In this 2-D scatter plot we can cluster datapoints of 4 different classes but still some datapoints are overlapped. So, neither of the 2-D scatter plots could give us significant classification outcomes. The graphs in Figure \ref{fig:eda_2} and Figure \ref{fig:eda_1}exhibits collinearity and to overcome this we extracted new features by performing feature engineering on current features by applying some mathematical functions like log, square, cube, etc. To classify the datapoints in the overlapping reqion was a big challenge. At the same time we had to consider time complexity, power and performance factors as well. This calls for more robust malware detection mechanism.

 
 

\begin{itemize}
    \item Here explain all the extracted features in detail - like one small paragraph each feature and how you extracted them
    
    \item Mention how you converted the binaries to the images.
    
    \item Any challenges you faced. Then how you extracted the features.
\end{itemize}

    
    

 
 
\section{Experimental Setup and Data Collection}
All the data collection was done using Python scripts for different classes of malware images and the extracted features. The experiments were run on a Windows 10 OS with Intel i7 Coffeelake, 32GB RAM, NVidia GTX 1080 Ti 8GB Graphics Card. The scripts were run in Jupyter notebook and the machine learning models were trained on the same machine with Weka tool \cite{Hall_weka'09}. 
\section{Proposed Mechanism to Detect Malware}
\label{proposed}

\begin{figure*}[!htb]
    \centering
    \includegraphics[width=0.9\textwidth]{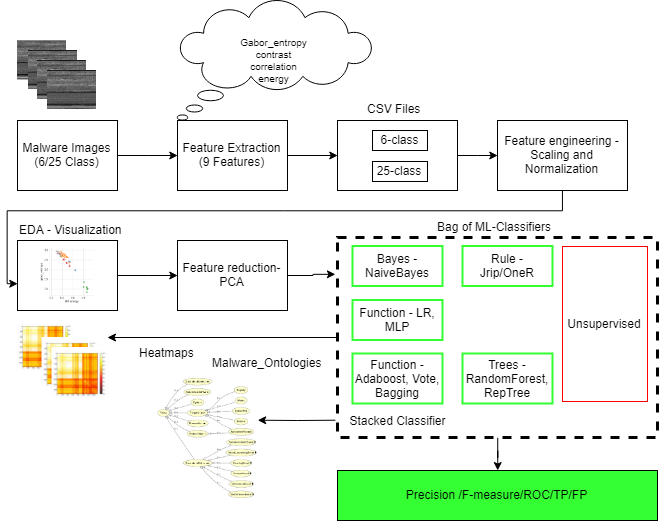}
    \vspace{1em}
    \caption{Dataset distribution with Classes and Sub-Classes}
    \label{fig:process_diagram}
\end{figure*}

Figure \ref{fig:process_diagram} depicts the entire process that we have performed to detect malwares and categorize them in different classes.The first three blocks of the Figure \ref{fig:process_diagram} namely Malware Images, Feature Extraction, constructing CSVs and EDA visualization have been discussed in Section \ref{background} in detail. Here we describe the other blocks of the diagram. 
So far we have learned all the background required for detecting malware using features corresponding to the malware images which were extracted from malware binaries as described in detail in Section \ref{background}. The conclusion of our EDA motivated that further analysis needed to be done to make fruitful use of the dataset created and process it in such a way that would give better and desired results after classification. The whole purpose is to prepare the data such that we do not need to tweak the classification models to a great extent and hence save a lot of time. 
Until now we are done with:

\begin{itemize}
    \item Obtaining the dataset
    \item Converting binaries to grayscale images
    \item Developing Python scripts to extract different features related to images from the dataset
    \item Generating CSV dataset files for the 6-main classes and 25-sub classes of malware
    \item Exploratory data analysis for dataset visualization of the 6-main and 25-sub classes to conclude what direction to proceed in
\end{itemize}

Depending upon the EDA results mentioned in Section \ref{background}, we were suggested to go for a feature selection technique that basically reduces our features to avoid irrelevant features being taken into consideration. 
\paragraph{\textbf{Feature Reduction and Selection using Principle Component Analysis}} There are many feature reduction and selection algorithms available but we decided to go with the most common PCA (Principle Component Analysis) algorithm. As the name suggests, PCA tries to minimize dataset to relevant features. There are many reasons why PCA component analysis is crucial before training classifiers:

\begin{itemize}
    \item Reducing Computation Time and Complexity: More the features, more time required to train the ML-model. Hence less complexity of the implemented model.
    \item Remove Irrelevant Features: There might be many features in the dataset that do not significantly contribute to 'relevant information' needed for best classification accuracy results, hence, removing or reducing the dataset to the most relevant ones will provide better accuracy under less time.
\end{itemize}

PCA is an algorithm that constructs totally new features known as Principle Components (PCs). These newly constructed features basically cover all the variance and information of the dataset thus reducing the complexity of models without compromising the information in the dataset. We had set the configuration in the Weka tool to include 95\% of the information of the dataset (both 6-class and 25-subclasses) with at the most 5 features while building the PC equations. The results of our PCA analysis is discussed in the Section \ref{results}. 

\begin{figure*}
    \centering
    \includegraphics[width=0.9\textwidth]{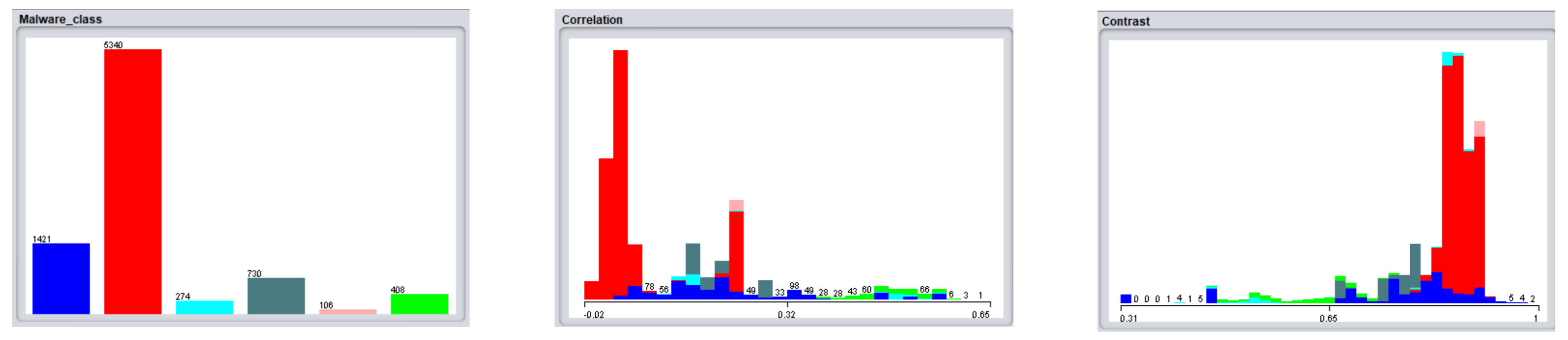}
    \vspace{1em}
    \caption{Visualization for 6-main classes; (a) All malware classes, (b) Correlation feature distribution and (c) Contrast feature distribution}
    \label{fig:visu_six}
\end{figure*}

\begin{figure*}
    \centering
    \includegraphics[width=0.9\textwidth]{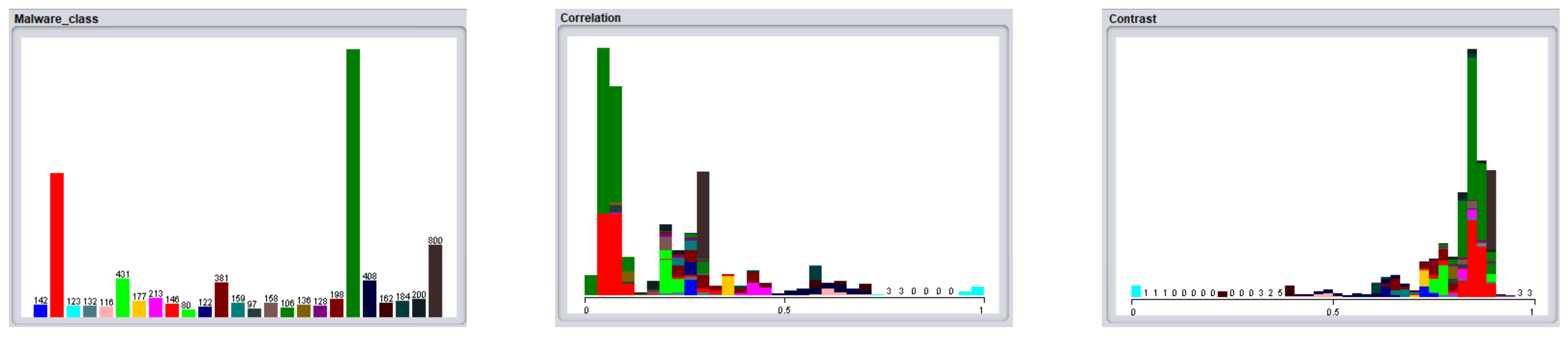}
    \vspace{1em}
    \caption{Visualization for 25-sub classes; (a) All malware sub-classes, (b) Correlation feature distribution and (c) Contrast feature distribution}
    \label{fig:visu_sub}
\end{figure*}

\paragraph{\textbf{Feature Engineering - Scaling and Normalization}} Usually the dataset that is built using the extracted features is not `clean' and needs a lot of preprocessing to ensure optimal detection accuracy of the classifiers and its correct functioning as well. In our case, since we built the feature extraction and CSV scripts such that it was preprocessed even before the data was saved to CSV files hence eliminating the need of `cleaning' the dataset. The only thing that we needed to do was data normalization which is a prerequisite of the PCA analysis. We normalized the dataset with the help of a filter available in the Weka tool named ``normalize". This scales and normalizes and entire columns in the dataset and also does the same across the columns. This process was much needed to ensure our classification stage does not take any performance hit. Figures \ref{fig:visu_six} and \ref{fig:visu_sub} show the visualization of the main and the sub-classes of malware dataset for correlation and contrast features. It shows how the data is distributed across various classes of malware and we have chosen only those graphs that have distribution spread across the x-axis, whereas all the other feature graphs were not so much spread and the columns overlapped each other - this was resolved when we trained the ML classifiers with the dataset because the problem is solved in higher dimensional space as against 2D visualization. 

\textbf{Bag of Classifiers} After the visualization was done on the dataset, we have trained our classifier(s) with the dataset. We have named this section as 'bag of ML classifiers' because we have trained multiple classifiers -each of a different type- to observe and draw conclusions as to which one performs the best. I our case, we have used the supervised type of classifiers which in our case have shown promising results as against unsupervised methods that rendered less than 30\% detection accuracy and needed further improvisations and hence we did not include those results here. We used the Weka tool \cite{Hall_weka'09} to perform all the ML training and testing. We have used the following types of classifiers which are unique in themselves:

\paragraph{\textbf{Naive Bayes}} Naive Bayes is a simple, yet effective and commonly-used, machine learning classifier. It is a probabilistic classifier that makes classifications using the Maximum A Posteriori decision rule in a Bayesian setting. It can also be represented using a very simple Bayesian network. Naive Bayes classifiers have been especially popular for text classification, and are a traditional solution for problems such as spam detection \cite{naive_bayes}.

\paragraph{\textbf{Random Forest}}
Random Forest is a flexible, easy to use machine learning algorithm that produces, even without hyper-parameter tuning, a great result most of the time. It is also one of the most used algorithms, because it’s simplicity and the fact that it can be used for both classification and regression tasks.Random Forest is a supervised learning algorithm. Random forest builds multiple decision trees and merges them together to get a more accurate and stable prediction \cite{random_forest}.

\paragraph{\textbf{Logistic Regression}} 
Logistic regression is a classification algorithm used to assign observations to a discrete set of classes. Unlike linear regression which outputs continuous number values, logistic regression transforms its output using the logistic sigmoid function to return a probability value which can then be mapped to two or more discrete classes. LR could help use predict whether the class under consideration is a rootkit or a worm. Logistic regression predictions are discrete - only specific values or classes are allowed \cite{logistic_regression}. All the results of the classifiers are presented in Section \ref{results}.

\begin{figure*}
    \centering
    \includegraphics[width=1\textwidth]{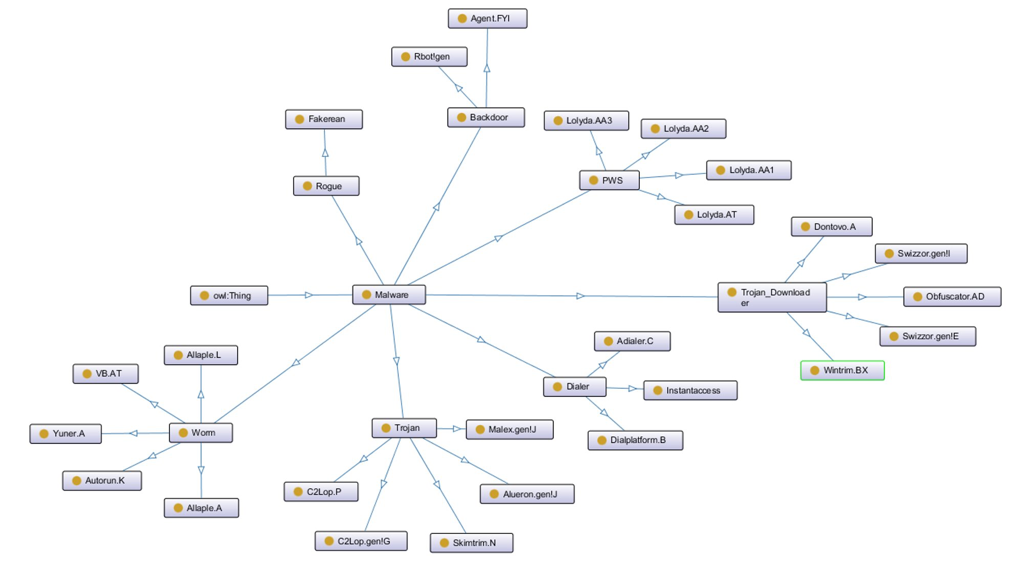}
    \vspace{10pt}
    \caption{Ontograph for malimg-dataset\vspace{10pt}}
    \vspace{1pt}
    \label{fig:ontograph}
\end{figure*}

\paragraph{\textbf{Ontograph with Protege Tool}}
Figure \ref{fig:ontograph} illustrates the ontograph for malimg dataset. We obtained this ontograph by using protege tool. Protege tool is a free, open source ontology editor and a knowledge management system. Protege provides a graphic user interface to define ontologies. It also includes deductive classifiers to validate that models are consistent and to infer new information based on the analysis of an ontology \cite{protege}.

\section{Results}
\label{results}
We here discuss and present all the results that we gathered for various sections of our project discussed so far. The results are presented in the same order as we have mentioned in the earlier sections. 

\subsection{PCA Analysis} The results of the 6-main class and 25-sub class PCA analysis is as shown in Tables \ref{tbl:six_class_pca} and Table \ref{tbl:sub_class_pca} respectively. The tables clearly indicate the features that were used to build the new principle components (PCs) and how much information each of them carries is mentioned in the 'attribute' column of the tables. These PCs can be used to feed and train the ML-classifier exploiting the benefits mentioned previously in Section \ref{proposed}. It is to be noted that we did not use the results of the PCA for training our classifiers because with the existing dataset we could get better accuracy and performance of the ML-models with limited time complexity and hence it was not needed to use PCA results but we still did include the results just to show that PCA is also a good option for achieving better accuracy with less time complexity. 

\begin{table*}[!htb]
\centering
\caption{PCA Results for 6-main class of malware}
\label{tbl:six_class_pca}
\scalebox{1}{
\begin{tabular}{ll}
\hline
Ranked & Attributes \\ 
\hline
\hline 
0.0808 & 1 -0.345Energy-0.342LBP energy-0.342Homogeneity+0.34 Dissimilarity+0.339LBP entropy \\
\hline
0.0289 & 2 -0.691Correlation+0.454gabor\_energy-0.375gabor\_entropy+0.227Dissimilarity-0.212Homogeneity \\
\hline
\hline
\end{tabular}
}
\end{table*}

\begin{table*}[!h]
\centering
\vspace{2em}
\caption{PCA Results for 25-sub class of malware}
\label{tbl:sub_class_pca}
\scalebox{1}{
\begin{tabular}{ll}
\hline
Ranked & Attributes \\ 
\hline
\hline 
0.7445 & 0.342Energy-0.341Homogeneity \\
\hline
0.6981 & 2 -0.567Malware\_class=Allaple.A-0.357Malware\_class=Lolyda.AA3 \\
\hline
0.6591 & 3 -0.776Malware\_class=Allaple.L+0.524Malware\_class=Allaple.A \\
\hline
0.6234 & 4 -0.681Malware\_class=Yuner.A-0.253Contrast \\

\hline
\hline
\end{tabular}}
\end{table*}

\subsection{Classification Results}
\begin{figure*}
    \centering
    \includegraphics[width=1\textwidth]{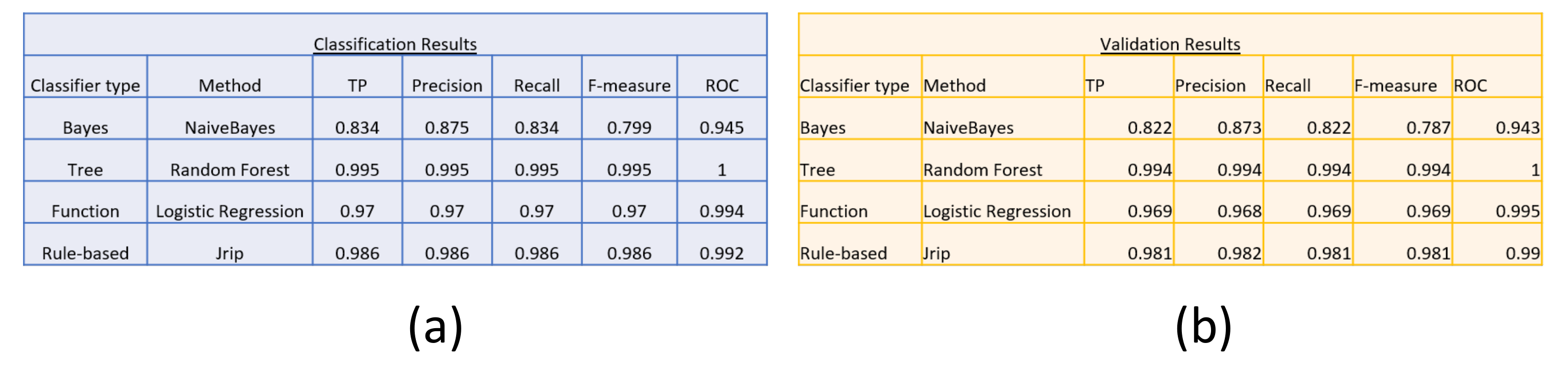}
    \vspace{1em}
    \caption{Results for 6-main classes (a) Classification results, (b) Validation Results}
    \label{fig:six_class_results}
\end{figure*}

\begin{figure*}
    \centering
    \includegraphics[width=1\textwidth]{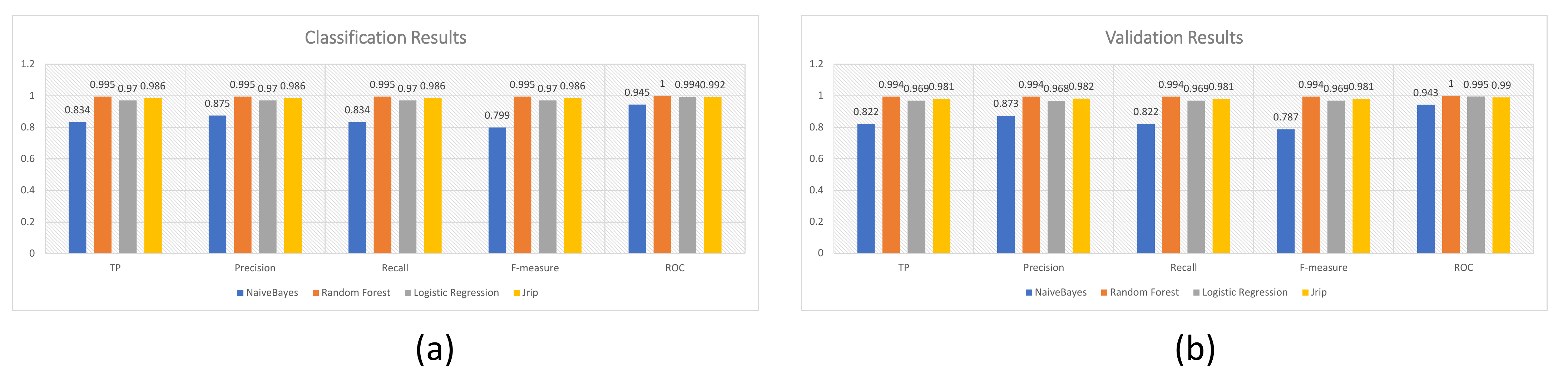}
    \vspace{1em}
    \caption{Results for 6-main classes (a) Classification bar graphs, (b) Validation bar graphs}
    \label{fig:six_class_plots}
\end{figure*}

\begin{figure*}
    \centering
    \includegraphics[width=0.95\textwidth]{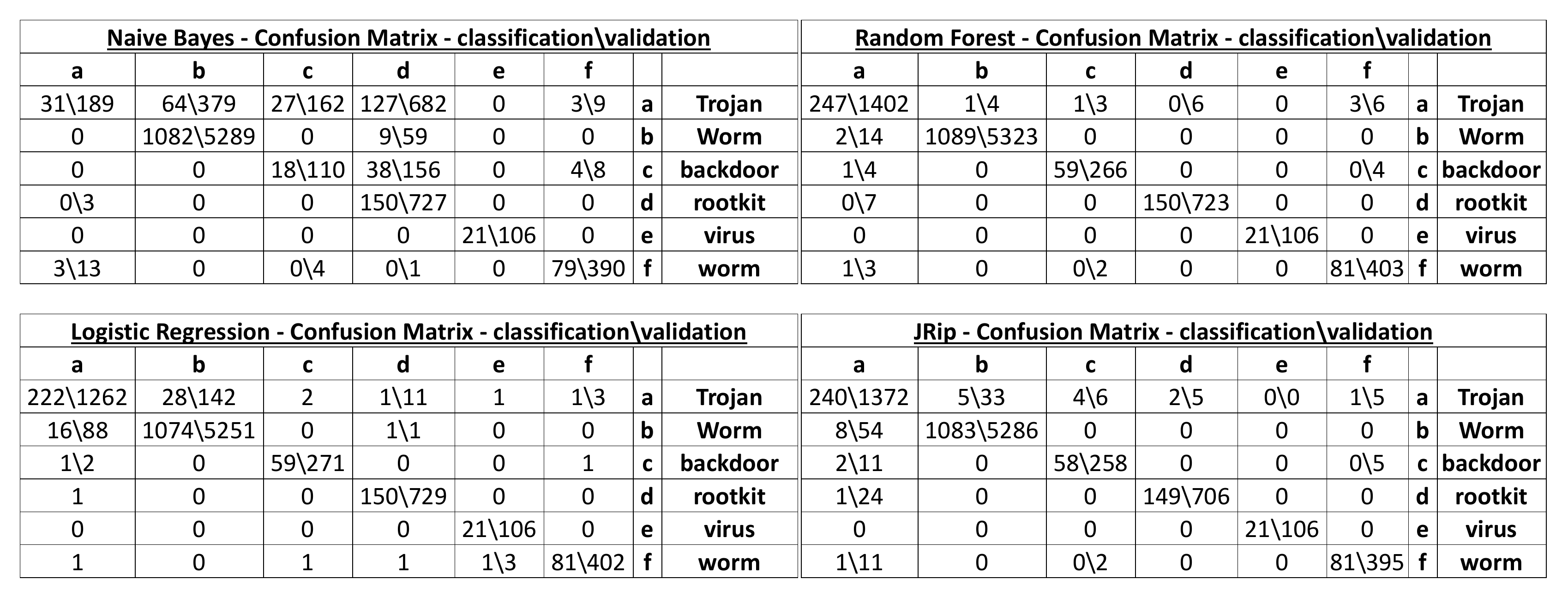}
    \vspace{1em}
    \caption{Confusion matrices for all the Classifiers for six main malware class}
    \label{fig:six_class_confusion}
\end{figure*}

\begin{figure*}
    \centering
    \includegraphics[width=0.95\textwidth]{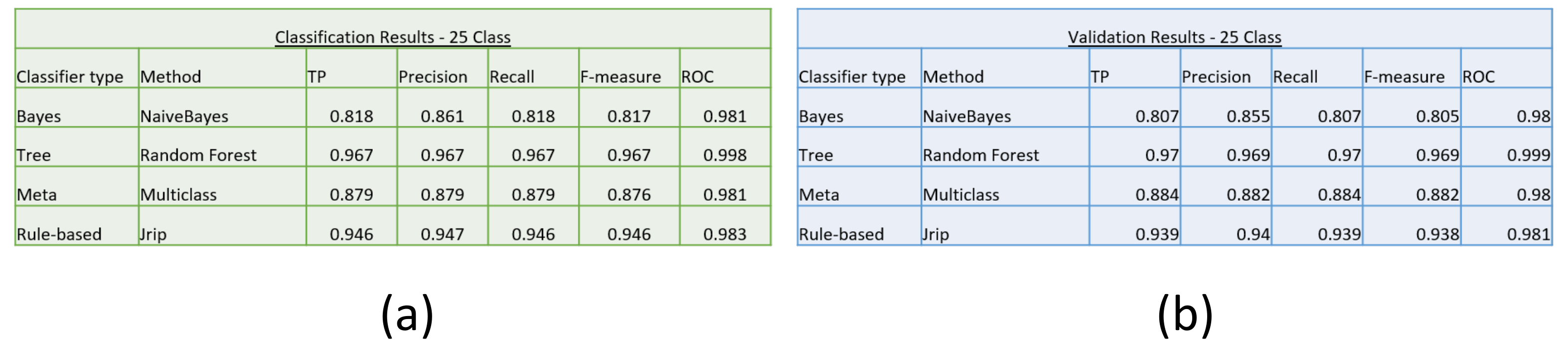}
    \vspace{1em}
    \caption{Results for 25-sub classes (a) Classification results, (b) Validation Results}
    \label{fig:two_five_results}
\end{figure*}

\begin{figure*}
    \centering
    \vspace{2em}
    \includegraphics[width=0.95\textwidth]{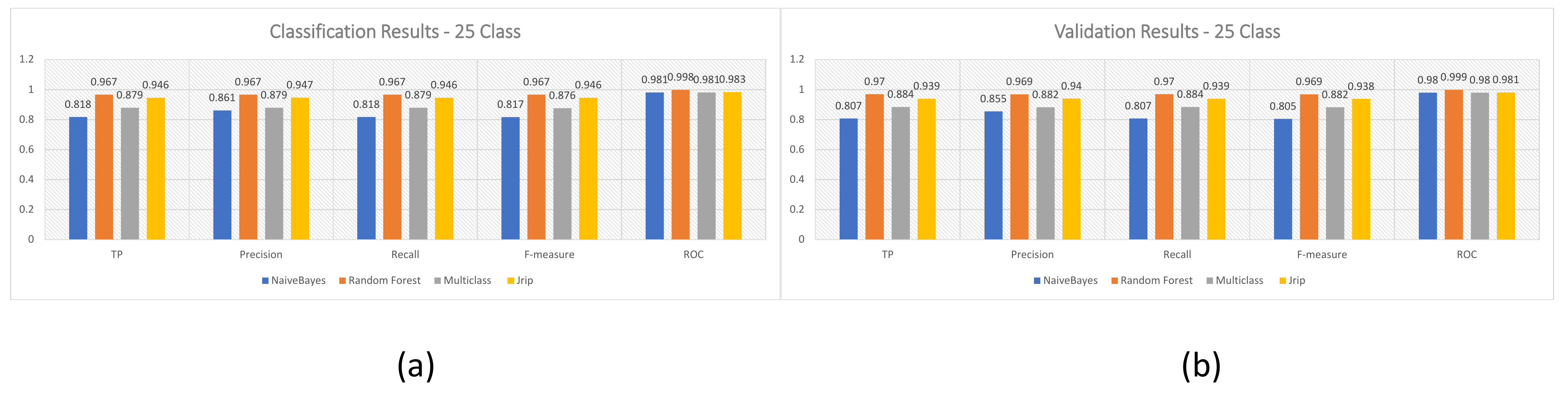}
    \vspace{1em}
    \caption{Results for 25-sub classes (a) Classification bar graphs, (b) Validation bar graphs}
    \label{fig:two_five_plots}
\end{figure*}

The results of all the classifiers that we have used are presented here with the conclusions that we can draw from them. Figure \ref{fig:six_class_results}(a),(b) show classification accuracy and other performance metrics as explained in Section \ref{appen}. Figure \ref{fig:six_class_results} shows the results of 6-main classes of malware after training and testing with 4-different machine learning models. We have used 80-20\% train/test while training the models. Figure \ref{fig:six_class_plots} shows the bar graph plots for the same data discussed earlier. The bar graphs show results with different performance metrics and we can conclude that all the classifiers used satisfy the given dataset and the problem of linear EDA graphs discussed in Section \ref{background} where all the plots were kind of linear were solved after classification results as the classification is done in higher dimensions. We did 2 fold validation on each classifier and did not go with higher fold validations as it took a lot of time to validate. The same results have been plotted for 25-sub classes of malware as shown in Figure \ref{fig:two_five_results} and Figure \ref{fig:two_five_plots}. These graphs show that with 25 class dataset, the results are better although some of the classes overlap which might be because two or more classes show similar characteristic features. 
Figure \ref{fig:six_class_confusion} shows the confusion matrix for 

\section{Related Works}
\label{sota}
This section discusses some of the previous works that have been published in the past that have proposed methods to detect malwares. Some of the techniques discussed here are software-based while others are hardware-based.
Works in \cite{sayadi-CF18,DAC4'17} extensively described how hardware performance counters (HPCs) can be used to detect anomalies in applications and classify malware as against benign employing ML-classifiers. The authors have used HPCs to feed to different set of classifiers and presented their results. They have also proposed malware detection for resource constrained systems where performance counters are limited and where systems resources have to used sparingly. Hence, they propose the use of Ensemble learning methods to boost the performance of general ML-classifiers.
\cite{Demme'13} has discussed about the feasibility of using HPCs for malware detection. They have also used ML-models to classify applications and supported their claim. In \cite{Sanket_ICMLA'19, Sanket_ICTAI'19} authors detect stealthy malwares by converting malware binaries into grayscale images and then extracting patterns by performing raster scanning. The grayscale images are further represented as sequence of patterns which are further used for sequence classification using RNN-LSTM's. Work in \cite{Sanket_CASES'19} introduces a hybrid approach which utilizes the microarchitectural traces obtained through on-chip embedded hardware performance counters (HPCs) and the application binary for malware detection. The obtained HPCs are fed to multi-stage machine learning (ML) classifier for detecting and classifying the malware. Authors in \cite{Sanket_DAC'21}  presents a collaborative machine learning (ML)-based malware detection framework. It introduces a) performance-aware precision-scaled federated learning (FL) to minimize the communication overheads with minimal device-level computations; and (2) a Robust and Active Protection with Intelligent Defense strategy against malicious activity (RAPID) at the device and network-level due to malware and other cyber-attacks. 
\cite{Rootkit-Singh} has proposed how kernel-level rootkits can be detected using HPCs on hardware level. They have described the process of training ML-models with the acquired HPCs and then presented results in support of their claim. They have tested their proposed mechanism by feeding the detector with both rootkit and clean traces. 
Authors in \cite{Bahador'14} interestingly used SVD (Singular-value decomposition) matrix in collaboration with HPCs to train ML-models to detect malwares. This is kind of partial software and partial hardware based approach in detecting anomalies. 
\cite{Jacob'08} has proposed a software based approach to detect malicious piece of code.
\section{Conclusion}
\label{conclusion}

We have first described the process of converting malware binaries to images and then explained the process of feature extraction. With the help of our results we have supported our claim that malware binary images can be used to detect classes of malware from each other with high accuracy and with a variety of classifiers. We then have presented the confusion matrices to show the classification efficiency of the employed models. Protege tool has been used to obtain ontograph for the dataset based on the main and sub classes. We hope to develop this method and take it further ahead in classifying malwares from benign and also improvise our time complexity by implementing this on hardware accelerated GPU. We also plan to improve our Ontology by adding `experiences' to malware instances. 

\begin{table}[!ht]
\centering
\vspace{2em}
\caption{accuracy, recall and precision matrix }
\label{tbl:f1-matrix}
\scalebox{1}{
\begin{tabular}{lll}
\hline
Actual/Predicted & Negative & Positive \\ 
\hline
\hline 
Negative & True Negative & False Positive \\
\hline
Positive & False Negative & True Positive \\
\hline
\hline
\end{tabular}
}
\end{table}
\section*{APPENDIX}
\label{appen}

For better understanding of the terms like accuracy, recall and precision refer to Table \ref{tbl:f1-matrix}. In this table first row belongs to predicted values and first column belongs to actual values. For example, if the actual value is negative and predicted value is negative then it is True negative. If actual value is negative and predicted value is positive then it is a false positive. Whereas if actual value is positive and predicted value is negative then it is false negative and if actual value is positive and predicted value is also positive then it is true positive.

So, from the above explanation the following are the equations for precision and recall,

Precision = (True positive)/ (True positive + False positive)

Recall = (True positive)/ (True positive + False negative )

\bibliographystyle{ieeetr}
\bibliography{references}

\end{document}